\documentclass[twoside,fleqn]{article}
\usepackage{espcrc2}
\usepackage{theorem}
\usepackage{epsfig}
\usepackage{graphicx}

\theoremstyle{break}    
\theoremstyle{plain}    
\theoremstyle{plain}    
\theoremstyle{plain}    
{\theorembodyfont{\rmfamily}     }
{\theorembodyfont{\rmfamily}     }

\def\gsim{{\mathrel{\raise2pt\hbox to 8pt{\raise -5pt\hbox{$\sim$}\hss{$>$}}}}}
\def\rsim{{\mathrel{\raise2pt\hbox to 8pt{\raise -5pt\hbox{$\sim$}\hss{$>$}}}}}
\def\lsim{{\mathrel{\raise2pt\hbox to 8pt{\raise -5pt\hbox{$\sim$}\hss{$<$}}}}}

\begin{document}

\title{
Calculating weak matrix elements using HYP staggered fermions\thanks{
Presented by W.~Lee.  Research supported in part by BK21, by the SNU
foundation \& Overhead Research fund, by KRF contract
KRF-2002-003-C00033, by KOSEF contract R01-2003-000-10229-0, by US-DOE
contract DE-FG03-96ER40956/A006 and by US-DOE contract W-7405-ENG-36.}
}
\author{T.~Bhattacharya\address[LANL]{MS--B285, T-8,
                Los Alamos National Lab, 
		Los Alamos, New Mexico 87545, USA},
        G.T.~Fleming\address{Jefferson Laboratory,
	        MS 12H2, 12000 Jefferson Avenue, Newport News,
	        VA 23606, USA},
        G.~Kilcup\address{Department of Physics,
                Ohio State University, Columbus, OH 43210, USA},
        R.~Gupta\addressmark[LANL],
        W.~Lee\address{School of Physics, 
		Seoul National University,
		Seoul, 151-747, South Korea}
        and
        S.~Sharpe\address{Department of Physics,
                University of Washington, 
		Seattle, WA 98195, USA}
}
\begin{abstract}
We present preliminary results of weak matrix elements relevant to
CP violation calculated using the HYP (II) staggered fermions.
Since the complete set of matching coefficients at the one-loop level
became available recently, we have constructed lattice operators with
all the $g^2$ corrections included.
The main results include both $\Delta I = 3/2$ and $\Delta I = 1/2$
contributions.
\end{abstract}

\maketitle


%
\section{Introduction}\label{sec:intr}
Staggered fermions preserve enough chiral symmetry to calculate the
weak matrix elements for CP violations ($B_K$, $\epsilon'/\epsilon$)
and have an advantage over DWF \cite{ref:CP-PACS:0,ref:RBC:0} of
requiring less computing, which makes dynamical simulations possible
below the physical strange quark mass.
In the previous attempt to calculate $\epsilon'/\epsilon$ using
unimproved staggered fermions \cite{ref:wlee:1}, we observed, in
$B_6$, a large dependence on the implementation chosen for the
quenched approximation as well as large perturbative corrections at
the one loop level.
In addition, it has been known that unimproved staggered fermions
have large scaling violations of order $a^2$ \cite{ref:jlqcd:1}.
Some of these problems can be alleviated by improving staggered
fermions using fat links.
In order to find the best improvement scheme to reduce perturbative
correction and taste symmetry breaking, we calculated explicitly one
loop matching coefficients for various improved staggered fermion
actions and operators \cite{ref:wlee:2}.
As a result of this study, it turned out that the Fat7
\cite{ref:orginos:0} and HYP \cite{ref:hasenfratz:0} fat links, after
a higher level of mean field improvement, lead to the greatest
reduction.
In addition, the key features of fat links of HYP type are
(i) that they are local in the sense of involving only gauge links 
contained in hypercubes connected to the original links;
(ii) that they lead to the largest reduction in taste symmetry 
breaking in the spectrum \cite{ref:hasenfratz:0}; and 
(iii) that the one-loop renormalization is identical
between the HYP (II) (perturbatively improved coefficients) and 
$\overline{\rm Fat7}$ (SU(3) projected Fat7) \cite{ref:wlee:3}.
Several useful properties of HYP and $\overline{\rm Fat7}$ links
are presented in \cite{ref:wlee:3}.
Recently we have calculated the current-current diagrams to obtain the
perturbative matching coefficients for the four-fermion operators
constructed using the HYP/$\overline{\rm Fat7}$ links
\cite{ref:wlee:4}.
We find that the perturbative corrections are reduced down to about
10\% level by using the HYP/$\overline{\rm Fat7}$ links.
We also have calculated the penguin diagrams and the results
are reported separately \cite{ref:wlee:5}.
These two calculations provides a complete set of matching formula to
calculate $\epsilon'/\epsilon$ using the HYP/$\overline{\rm Fat7}$
staggered fermions.
Here, we will present preliminary estimates of $B_K$, $B_7^{(3/2)}$,
$B_8^{(3/2)}$ and $B_6^{(1/2)}$ calculated using the HYP (II)
staggered fermions at $\beta = 6.0$ on a $16^3 \times 64$ lattice with
218 configurations.
\section{$B_K$}
\begin{figure}[t]
\epsfig{file=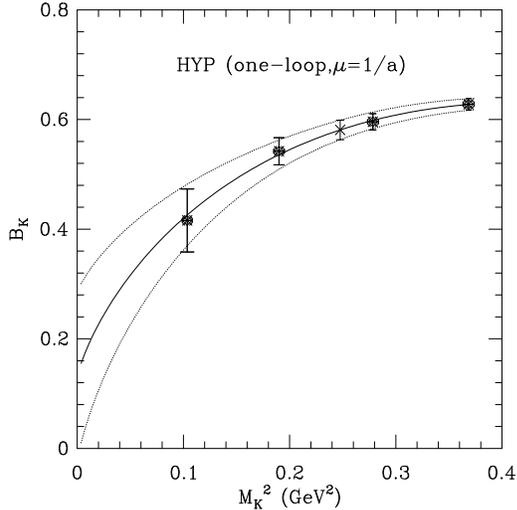, height=16pc, width=16pc}
\vspace*{-5mm}
\caption{$B_K(\mu=1/a)$}
\label{fig:b_k}
\vspace*{-3mm}
\end{figure}
Fig. \ref{fig:b_k} shows $B_K$ as a function of $M_K^2$, where the
mesons are made of degenerate quarks.  
We fit $B_K$ to the form suggested by (quenched) chiral perturbation
theory: $B_K = c_0 + c_1 (M_K)^2 + c_2 (M_K)^2 \log(M_K)^2$.
The cross symbol in Fig. 1 corresponds to the value interpolated to
the physical kaon mass.
Our preliminary result is $B_K = 0.581(18)$, which is consistent with
the continuum extrapolated value calculated using unimproved staggered
fermions \cite{ref:jlqcd:1}.
In the chiral limit, we obtain $c_0 = 0.13(15)$, which is also
consistent with those results obtained using the NLO, large $N_c$
calculation \cite{ref:peris:1}.
The value for $c_2/c_0$ is consistent with the predictions of quenched
chiral perturbation theory \cite{ref:sharpe:1} within large errors.
\section{$B_7^{(3/2)}$ and $B_8^{(3/2)}$}
\begin{figure}[t]
\epsfig{file=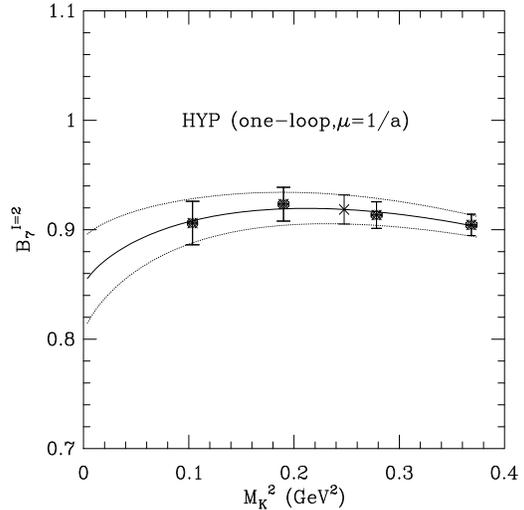, height=16pc, width=16pc}
\vspace*{-5mm}
\caption{$B_7^{\Delta I=3/2}(\mu=1/a)$}
\label{fig:b_7}
\vspace*{-3mm}
\end{figure}
\begin{figure}[t]
\epsfig{file=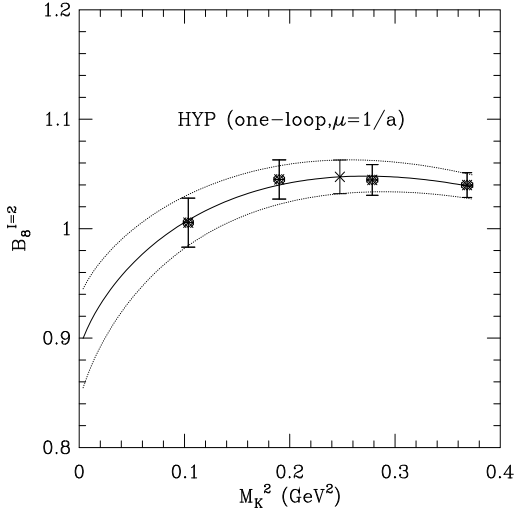, height=16pc, width=16pc}
\vspace*{-5mm}
\caption{$B_8^{\Delta I=3/2}(\mu=1/a)$}
\label{fig:b_8}
\vspace*{-3mm}
\end{figure}
A major contribution to the $\Delta I = 3/2$ amplitudes comes from
$B_7^{(3/2)}$ and $B_8^{(3/2)}$
The results are presented in Fig.~\ref{fig:b_7} and Fig.~\ref{fig:b_8}.
We fit $B_7$ and $B_8$ to the form suggested by chiral
perturbation theory: $B_{7,8} = c_0 + c_1 (M_K)^2 + c_2 (M_K)^2
\log(M_K)^2$.
Our preliminary values at the physical kaon mass are 
$B_7^{(3/2)}=0.919(13)$ and $B_8^{(3/2)}=1.047(15)$.
Note that we calculated $B_7^{(3/2)}$ at the scale $\mu = 1/a$ using
the HYP staggered fermions, which would not have been meaningful for
unimproved staggered fermions due to large perturbative corrections.
Compared with previous calculation done using Landau-gauge operators
\cite{ref:kilcup:1}, the systematics of the HYP staggered operators
are significantly reduced and the results are more reliable.
\section{$B_6^{(1/2)}$}
\begin{figure}[t]
\epsfig{file=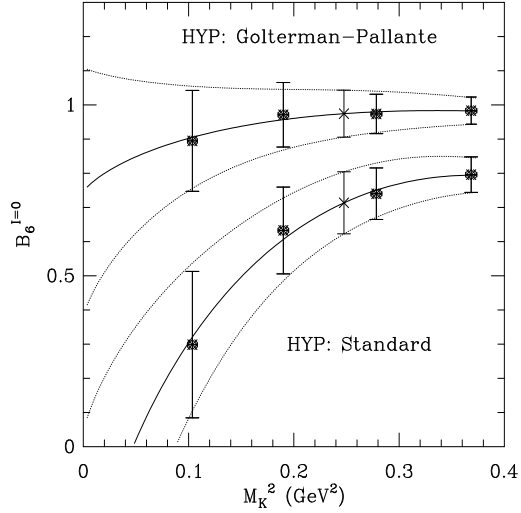, height=16pc, width=16pc}
\vspace*{-5mm}
\caption{$B_6^{\Delta I=1/2}(\mu=1/a)$}
\label{fig:b_6}
\vspace*{-3mm}
\end{figure}
%
%
%
A major contribution to $\Delta I = 1/2$ amplitudes comes from
$B_6^{(1/2)}$.
There are two independent methods to calculate $B_6^{(1/2)}$: the
standard (STD) method and the Golterman-Pallante (GP) method
\cite{ref:golterman:1}.
Fig.~\ref{fig:b_6} shows $B_6^{(1/2)}$ as a function of $M_K^2$.
Note that unlike the unimproved staggered fermion calculations where
the perturbative corrections to the STD calculation are $\approx
50\%$, the perturbative corrections in this calculation are modest for
both the STD and GP methods.
We also observe that the gap between the STD and GP methods is
reduced at the physical kaon mass using the HYP staggered fermions
compared that of the unimproved staggered fermions.
We fit $B_6$ to the form suggested by chiral perturbation theory: $B_6
= c_0 + c_1 (M_K)^2 + c_2 (M_K)^2 \log(M_K)^2$.
Our preliminary values at the physical kaon mass are
\begin{eqnarray*}
B_6^{(1/2),STD}(\mu=1/a) &=& 0.714(91) \\
B_6^{(1/2),GP}(\mu=1/a) &=& 0.974(69) \,.
\end{eqnarray*}
\section{Preliminary $\epsilon'/\epsilon$}
We use the formula given in \cite{ref:buras:1} to convert
$B_6^{(1/2)}$ and $B_8^{(3/2)}$ into $\epsilon'/\epsilon$.
When we use the STD method for $B_6^{(1/2)}$, $\epsilon'/\epsilon
(STD) = 0.00046(23)$.
For the GP method for $B_6^{(1/2)}$, $\epsilon'/\epsilon (GP) =
0.00115(17)$.
These values are very preliminary and we have not included an analysis
of the systematic errors.
In addition, we did not use lattice values for any $B_i^{(1/2)}$
except for $B_6^{(1/2)}$, since we have not yet extracted them.
Hence, we plan to calculate all the $B_{i}^{(1/2)}$ and incorporate
all of them into the calculation of $\epsilon'/\epsilon$.
We also plan to obtain the optimal matching scale, $q^*$
\cite{ref:wlee:6}.
We plan to extend our calculation to dynamical simulation
using the HYP staggered fermions.
We thank N.~Christ, C.~Jung, C.~Kim, G.~Liu, R.~Mawhinney and L.~Wu
for their support of this project and assistance with numerical
simulations on the Columbia QCDSP supercomputer.
%

%
%
%
%
%
%
%
%
%
%
%

%
%

\end{document}